\documentclass[twocolumn,aps,prd,amsfonts,amsmath,amssymb,nofootinbib,longbibliography,preprintnumbers]{revtex4-1}

\usepackage{ifpdf}

\ifpdf
    \usepackage[pdftex,colorlinks=true,plainpages=false,pdfpagelabels]{hyperref}
    \usepackage[pdftex]{color}
    \hypersetup{colorlinks = true}
\else
    \usepackage[colorlinks=true]{hyperref}
\fi

\usepackage[hmargin=1.5cm,vmargin=1.75cm]{geometry}

\usepackage{xspace}
\newcommand{\cp}[1]{\ensuremath{\mathbb{CP}^#1}\xspace}
\newcommand{\cpk}{\cp{k}}
\newcommand{\half}{\frac{1}{2}}
\newcommand{\diag}{\mathrm{diag}}

\def\qHOne{q_H}

\begin{document}

\preprint{IPMU13-0238}

\title{Charge Quantization and the Standard Model\\
       from the\\
       \cp2 and \cp3 Nonlinear $\sigma$-Models}

\author{Simeon Hellerman}
\email{simeon.hellerman.1@gmail.com}
\author{John Kehayias}
\email{john.kehayias@ipmu.jp}
\author{Tsutomu T.~Yanagida}
\email{tsutomu.tyanagida@ipmu.jp}

\affiliation{Kavli Institute for the Physics and Mathematics of the Universe (WPI)\\
             Todai Institutes for Advanced Study, The University of Tokyo\\
             Kashiwa, Chiba 277-8582, Japan}

\begin{abstract}

  \noindent
  We investigate charge quantization in the Standard Model (SM)
  through a \cp2 nonlinear sigma model (NLSM), $SU(3)_G/(SU(2)_H
  \times U(1)_H)$, and a \cp3 model, $SU(4)_G/(SU(3)_H \times
  U(1)_H)$. We also generalize to any \cpk model. Charge quantization
  follows from the consistency and dynamics of the NLSM, without a
  monopole or Grand Unified Theory, as shown in our earlier work on
  the \cp1 model
  (\href{http://arxiv.org/abs/1309.0692}{arXiv:1309.0692}). We find
  that representations of the matter fields under the unbroken
  non-abelian subgroup dictate their charge quantization under the
  $U(1)_H$ factor. In the \cp2 model the unbroken group is identified
  with the weak and hypercharge groups of the SM, and the
  Nambu-Goldstone boson (NGB) has the quantum numbers of a SM
  Higgs. There is the intriguing possibility of a connection with the
  vanishing of the Higgs self-coupling at the Planck
  scale. Interestingly, with some minor assumptions (no vector-like
  matter and minimal representations) and starting with a single quark
  doublet, anomaly cancellation requires the matter structure of a
  generation in the SM. Similar analysis holds in the \cp3 model, with
  the unbroken group identified with QCD and hypercharge, and the NGB
  having the up quark as a partner in a supersymmetric model. This can
  motivate solving the strong CP problem with a vanishing up quark
  mass.

\end{abstract}

\maketitle

\section{Introduction}

The quantization of electric charge was observed many decades ago, and
remains an exquisitely confirmed aspect of nature today with no known
exception. This experimental fact has inspired several endeavors to
explain this mystery, the most well-studied and successful being the
Dirac monopole \cite{Dirac:1931kp} and Grand Unified Theories (GUTs)
beginning with Georgi and Glashow \cite{Georgi:1974sy}. Perhaps most
particle physicists' view is that the latter is mechanism is the more
relevant one: at some energy scale much higher than the reach of
current experiments, the gauge groups of the Standard Model (SM) are
unified, and electromagnetic charge quantization follows from this
unification into a single gauge group.

There are also several well-known drawbacks to GUTs and
monopoles. Monopoles in these theories tend to be very heavy and cause
cosmological problems, GUTs generically predict too fast a rate for
proton decay, splitting the Higgs doublet and triplet masses is
difficult, and so far no direct experimental evidence has been
found. These are old problems which have a host of proposed solutions
such as inflation, discrete symmetries, high mass scales, and so
on. While GUTs remain relevant for model-building and phenomenology,
it can be fruitful to think outside of the box (of GUTs).

In a previous work \cite{cp1paper}, we considered charge quantization
in a $\mathbb{CP}^1$, or $SU(2)_G/U(1)_H$, nonlinear sigma model
(NLSM). The subscripts $G$ and $H$ differentiate between an
approximate global symmetry and an unbroken subgroup (which will be
gauged and identified with some subset of the SM groups),
respectively. We found that charge is quantized in half-integer units
of the Nambu-Goldstone boson (NGB) charge. A key point is that the
$SU(2)_G$ is \textit{never} gauged --- it is only an approximate,
nonlinearly realized symmetry. The $U(1)_H$ is gauged and identified
with the $U(1)_Y$ hypercharge of the SM.\footnote{This gauging
  explicitly breaks the $SU(2)_G$, but this does not affect charge
  quantization. We require the presence of a consistent theory in the
  limit of Yukawa and gauge couplings vanishing. This leads to charge
  quantization, and as long as charge is conserved, these couplings
  cannot break charge quantization. This was addressed in
  \cite{cp1paper}, but we will comment more on this issue in
  subsequent work.} This model achieves charge quantization in the SM
without a monopole or in the context a GUT, avoiding all of the
associated problems. Furthermore, the NGB of this model is completely
stable and fractionally charged, with a mass that can be light with
intriguing phenomenological possibilities, such as applications in
nuclear physics or as dark matter.

The derivation of charge quantization in \cite{cp1paper} is
reminiscent\footnote{There is also some similarity to earlier work in
  theories with Wess-Zumino terms \cite{Rabinovici:1984mj}.} of the
arguments given via monopoles (in the modern understanding due to Wu
and Yang \cite{Wu:1975vq,*Wu:1976ge}), as we are requiring
well-defined transformation laws for a matter field over a sphere
(\cp1 as a manifold), or GUTs, as it is the group structure which
plays a critical role. However, our derivation is also rather
distinct: there is \textit{no monopole} and the $SU(2)_G$ is never
gauged nor linearly realized. We work directly with the NLSM as we
consider that this does not always imply the presence of a
corresponding linear model. One can think of the quantization
condition arising due to the compact origin of the unbroken $U(1)_H$,
and thus topological in nature. In this spirit, it is natural to
consider other NLSMs which have a similar origin for a $U(1)_H$
subgroup: we will see that charge is quantized here as well.

A seemingly unrelated question turns out to be intimately linked to
charge quantization in these models: why is the matter content of a
generation in the SM what it is? There is no a priori reason for the
structure of matter we observe, nor any relation between their quantum
numbers. In a GUT, we have complete representations of the GUT group,
and after breaking to the SM we have certain representations of the SM
gauge groups. As we will see, this NLSM realization of charge
quantization has a deep structure: it imposes a relation between the
charge and the representation of the unbroken nonabelian group. Such a
relation does not follow from a GUT construction, and in a NLSM we
are not bound to having complete multiplets of a linearly realized and
spontaneously broken group. This is a surprising and unique prediction
of this theory, which will also lead to the matter representations
comprising a generation of the SM.

In this work we will extend and generalize this technique for charge
quantization to other models, as well as discussing phenomenological
applications and other theoretical aspects. In this work we will
primarily be concerned with \cp2 (and \cp3), but this immediately
gives us charge quantization for general \cpk.  We begin by
summarizing the previous work on the \cp1 model and extending to \cp2
and \cpk models in Sec.~\ref{sec:cpk}. The phenomenology of these
models is explored in Sec.~\ref{sec:cpk_pheno}, where we see that the
\cp2 model has NGBs with the quantum numbers of a SM Higgs and with
minor assumptions we are lead naturally to the matter content of a
generation. The \cp3 model has a similar motivation for the SM
generation content in its supersymmetric extension, and the model can
link the NGB to a vanishing mass for the up quark and the strong CP
problem. Finally, we discuss further extensions and related topics and
give concluding remarks in Sec.~\ref{sec:conc}.

\section{Charge Quantization in \cpk Models}
\label{sec:cpk}

\subsection{Review of $\mathbb{CP}^1$}
\label{sec:cp1}
We start by briefly reviewing our earlier work on \cp1
models\footnote{In our earlier work we discuss why we consider a
  supersymmetric model, despite not needing supersymmetry directly in
  our derivation. To summarize, supersymmetry ensures that the
  K\"ahler structure of the model is protected once matter is
  added. We will not comment further about supersymmetry in this
  work.} (see \cite{cp1paper} for the full derivation) which will be
straightforwardly extended to the larger \cpk models. First, let us
define our coordinates for \cp1 as $\phi_{1,2}$ which satisfy the
defining property of \cp1, $(\phi_1, \phi_2) = (\lambda\phi_1,
\lambda\phi_2)$. The ratio of these coordinates (automatically
satisfying this property) are the affine coordinates $z_+ \equiv
v\phi_1/\phi_2$ and $z_- \equiv v^2/z_+$, where $v$ is the symmetry
breaking vev. In more physical terms (we are basically following the
construction of \cite{Coleman:1969sm, *Callan:1969sn}), the
$z$-coordinates are the Nambu-Goldstone modes of the breaking from
$SU(2)_G$ to $U(1)_H$, the NLSM description of \cp1.

The $SU(2)_G$ is nonlinearly realized, while the $U(1)_H$ is a good
(linearly realized) symmetry. By explicitly considering consistent and
well-defined transformation properties of a charged matter field, the
complex scalar (for simplicity, or any other type) field $\chi$, over
all of \cp1 we are led to a charge quantization condition.\footnote{If
  all of the fields in the NLSM form parts of complete linear
  multiplets of $G$, charge quantization follows trivially.}

The infinitesimal generators\footnote{Here we will only work
  explicitly with the holomorphic generators and to linear order in
  the scalar field. For a complete discussion, see \cite{cp1paper}.}
of $SU(2)_G$ are labeled as $T_+, T_-$ and $T_0$. \cp1, thought of as
the manifold $S^2$, needs two coordinate patches, which we call the
southern hemisphere ($z_- \neq 0$ everywhere, $z_+ = 0$ at the south
pole) and the northern hemisphere ($z_+ \neq 0$ everywhere, $z_- = 0$
at the north pole).

Working first in the southern hemisphere with $z_+$ and $\chi$, the
action on $z_+$ is
\begin{subequations}
\begin{align}
\delta_{T_+}\circ z_+ &= -\frac{1}{v} z_+^2 \ ,\\
\delta_{T_-}\circ z_+ &= v\ ,\\
\delta_{T_0}\circ z_+ &= + z_+ \ .\label{normA}
\end{align}
\end{subequations}
The $U(1)_H$ charge is defined as the eigenvalue under $T_0$, with the
NGB $z_+$ having charge $+1$. $\chi$ has charge $\alpha$ and a
nonlinear transformation under the other (broken) generators of
$SU(2)_G$. After using the $SU(2)_G$ algebra and demanding the
transformations are smooth at the south pole ($z_+ = 0$) the
transformations on functions of $\chi$ and $z_+$ are determined to be
\begin{subequations}
\begin{align}
\delta_{T_+} &=  -\frac{2\alpha}{v} z_+\chi\partial_\chi - \frac{1}{v} z_+^2\partial_{z_+}\ ,\label{MatterTransA}\\
\delta_{T_-} &=  v\partial_{z_+}\ ,\label{MatterTransB}\\
\delta_{T_0} &= \alpha\chi\partial_\chi + z_+ \partial_{z_+}\ .\label{MatterTrans}
\end{align}
\end{subequations}

Switching to the northern hemisphere, we change coordinates to $z_-$. $\chi$ must also transform:
\begin{equation}
  \label{eq:chip}
  \chi' \propto z_-^{-p}\chi\ ,
\end{equation}
with the form fixed by $\chi$ and $\chi'$ having definite eigenvalues under
the same $U(1)_H$ --- antipodal points are fixed by the same rotation
generator, therefore the unbroken $U(1)_H$ at the two poles can be
identified. Performing the coordinate and field transformations the
full generators in the northern hemisphere are
\begin{subequations}
\begin{align}
\delta_{T_0} &= - z_- \partial_{z_-} + \left(\alpha + p\right) \chi'\partial_{\chi'}\ ,\\
\delta_{T_-} &= -\frac{z_-}{v} \left( z_-\partial_{z_-} - p \chi'\partial_{\chi'}\right)\ ,\\
\delta_{T_+} &=  v\partial{z_-} - v\left(p+ 2\alpha\right) z_-^{-1} \chi'\partial_{\chi'}\ .
\end{align}
\end{subequations}

Requiring that the transformations be well-defined at the north pole,
$z_- = 0$, and that the transformation to $\chi'$ is single-valued
everywhere forces
\begin{equation}
  \label{eq:cp1cq}
  p = -2\alpha \in \mathbb{Z}.
\end{equation}
Therefore the charge $\alpha$ of the matter field $\chi$ is quantized in
half-integer units of the NGB charge ($+1$ for $z_+$).

Another way to derive this charge quantization condition is to
consider the kinetic terms for a charged field.\footnote{The NGB's
  kinetic terms are completely fixed by the \cp1 structure and the
  Fubini-Study metric.} The K\"ahler potential is fixed by enforcing
that it is invariant under the full holomorphic plus antiholomorphic
$SU(2)_G$ transformations, as well as being quadratic in $\chi,\chi^\dagger$, and
invariant under phase rotations. It is given by
\begin{equation}
  \label{eq:kmatter}
  K_\mathrm{matter} = \left(1 + \frac{|z_+|^2}{v^2}\right)^{-2\alpha}|\chi|^2
\end{equation}
in the southern hemisphere. Requiring that the kinetic terms have the
same well-defined form in the northern hemisphere of \cp1 leads to the
same quantization condition given above.

\subsection{$\mathbb{CP}^k$ Models}
\label{sec:cpkmodels}
Although such an explicit derivation should be possible, in principle,
for the larger \cpk models, it quickly becomes rather unwieldy. Now
that we have a detailed understanding of the \cp1 model, we can
exploit it to understand the general \cpk models. We will do this by
adding mass terms for the ``extra'' (beyond \cp1) NGBs and flowing
through renormalization to the \cp1 model. In this way we will be able
to derive a general charge quantization formula for any \cpk model.

Let us do this explicitly for the \cp2 model and use this to
generalize to larger $k$. We label the NGBs as $(z_1, z_2)$ from the
group breaking of $G = SU(3)_G$ to $H = SU(2)_H \times U(1)_H$. The
NGBs transform as fundamentals of the unbroken $SU(2)_H$. The
generator of $U(1)_H$ is $\qHOne$ normalized by acting on the NGBs
with charge $+1$. We will consider a matter field $\chi$ which is
coupled in a $SU(3)_G$-symmetric way. $\chi$ is labeled by its
representation/charge under the unbroken subgroup; the representations
cannot necessarily be chosen arbitrarily.

We will label the eigenvalue of $\chi$ under $U(1)_H$ as $q_\chi$. The
representation of $\chi$ under $SU(2)_H$ is given by the ``2-ality''
$T$ (i.e.~``$k$-ality'' with $k = 2$) of the representation: its
eigenvalue, $\pm 1$, under the center of the group. With $J = T/2$ the
isospin of the $SU(2)_H$ representation, we have $(-1)^T =
(-1)^{2J}$. An important point is that the central element, $C$, can
be generated by exponentiating any \textit{non}-central element. If
$\hat{{\bf K}}_2$ is any generator of $SU(2)_H$ normalized to have
eigenvalues $\pm 1/2$ in the fundamental representation, then $C =
\exp(2\pi i\hat{{\bf K}}_2)$. Comparing with $C = \exp(\pi i T)$, we
see that $2\hat{{\bf K}}_2$ measures the ``2-ality'' $T$, defined
modulo $2$. For convenience, we will take $\hat{{\bf K}}_2$ to be the
element that acts on the NGB doublet $(z_1, z_2)$ as the diagonal
matrix with eigenvalues $\pm 1/2$.

Let us now add a mass term for one of the NGBs, $z_1$. We can do this
while preserving an $SU(2)$ subgroup of $G$, which we call $G' \equiv
SU(2)_{G'}$ (note: this is \textit{not} the same subgroup as
$SU(2)_H$). The theory then flows to the \cp1 NLSM, where the unbroken
subgroup is $U(1)_{H'}$.

After the renormalization group flow, the matter field $\chi$, which was
coupled to the \cp2 model in a $SU(3)_G$ symmetric way, is now coupled
to the \cp1 NLSM a way which preserves $SU(2)_{G'}$. This is by virtue
of the fact that the $SU(2)_{G'}$ is a subgroup of $SU(3)_G$ that is
preserved by the mass term. We can now apply the charge quantization
condition we derived for the \cp1 model: $q_{H_1'} = n/2$, where $n \in
\mathbb{Z}$ and $q_{H_1'}$ is the eigenvalue for $\chi$ under $U(1)_{H'}$.

We now want to relate this charge to the eigenvalues for the matter
field in the original \cp2 NLSM. We exploit the fact that there must
be a linear relation among $\hat{{\bf K}}_2, \qHOne,$ and $q_{H'}$ as
these are $3$ commuting generators of $SU(3)_G$, which is only rank
$2$. With the $SU(3)_G$ generators as $t^A_{SU(3)_G}$, the Pauli
matrices labeled $\sigma^a_{SU(2)}$, and writing the other group
generators in (block) diagonal form as
\begin{align*}
  \label{eq:cp2gen}
  SU(3)_G = t^A_{SU(3)}, \quad & SU(2)_{H} = \diag\{\sigma^a_{SU(2)}, 0\},\\
  \hat{{\bf K}}_2 = \diag\left\{+\half, -\half, 0\right\}, \quad & \qHOne = \diag\left\{+\frac{1}{3}, +\frac{1}{3}, -\frac{2}{3}\right\},\\
  SU(2)_{G'} = \diag\{0, \sigma^a_{SU(2)}\}, \quad & q_{H'} = \diag\left\{0, +\half, -\half\right\},
\end{align*}
the relation between the generators is
\begin{equation}
  \label{eq:linrel}
  \qHOne = \frac{4}{3}q_{H'} + \frac{2}{3}\hat{{\bf K}}_2.
\end{equation}
Using the known charge quantization condition in \cp1 and rewriting in
terms of the ``2-ality'' of the $SU(2)_G$ representation of $\chi$, we
have a charge quantization for a matter field in the \cp2 model, with
$n \in \mathbb{Z}$:
\begin{equation}
  \label{eq:cqcp2}
  q_\chi = \frac{2n}{3} + \frac{1}{3}(``2\mathrm{-ality}" \bmod{2}~\mathrm{of}~\chi).
\end{equation}
More explicitly, the quantization condition can be written as
\begin{equation}
  \label{eq:cp2cq2}
q_\chi = \begin{cases}
 \frac{2n}{3}, & \chi \mbox{ is a tensor respresentation of } SU(2)_H\\
 \frac{2n + 1}{3}, &  \chi \mbox{ is a spinor respresentation of }SU(2)_H
\end{cases}
\end{equation}
relative to the NGB charge ($+1$ in our conventions).

Having the charge quantization relation for both \cp1 and \cp2, we can
now see quite easily how this will generalize for arbitrary \cpk. We
add mass terms for all but one of the NGBs, preserving an $SU(2)_{G'}$
subgroup and flowing through the renormalization group to the \cp1
model. Let $T \pmod{k}$ represent the ``$k$-ality'' of the
representation of $\chi$ under the $SU(k)_H$. For example, in the \cp3
model $T = 0, -1, +1$ for a singlet, anti-fundamental, and fundamental
representation, respectively. The general charge quantization
condition is
\begin{equation}
  \label{eq:cpkcq}
  q_\chi = \frac{kn + (T \bmod{k})}{k + 1},
\end{equation}
relative to the NGB charge, which we always define as $+1$. The NGBs
are always in the fundamental representation of the unbroken
$SU(k)_H$. An interesting observation is that any non-singlet under
$SU(k)_H$ \emph{must} have nonzero $U(1)_H$ charge ($n \in \mathbb{Z}$
and $|T| < k$).

\section{Phenomenology}
\label{sec:cpk_pheno}

For phenomenology, and to successfully quantize electromagnetic
charge, we must relate the NLSM $U(1)_H$ and the $U(1)_Y$ hypercharge
of the SM. In order to fix the coefficient of proportionality between
these generators, we will use a ``minimality'' condition: the smallest
possible hypercharge should be the smallest hypercharge in the SM,
$1/6$.

In the \cp1 model (see \cite{cp1paper}) this led to hypercharges given
by $q_Y = n/6$. The NGB has a fractional charge, is exactly stable,
and has an electromagnetic mass from the gauging of $U(1)_Y$. This
particle can have a collider accessible mass, and has implications for
nuclear physics, especially nuclear fusion reactors. The NGB can be a
component or possibly all of the dark matter, depending on the
mass. We discussed the phenomenology of this model in more detail in
\cite{cp1paper}.

For the \cp2 model, the unbroken group is $SU(2)_H \times U(1)_H$. We want
to identify this with the SM weak group, $SU(2)_L$, and the
hypercharge group, $U(1)_Y$. The full SM is then $SU(3)_\mathrm{QCD}
\times \cp2$. All of the charges in eq.~\eqref{eq:cp2cq2} are given
relative to the NGB charge, normalized to $1$ in our conventions. In
this model the NGB has the quantum numbers of the SM Higgs
field. Thus, it is very interesting to identify the NGB with the SM
Higgs boson, setting the coefficient of proportionality between
hypercharge and $U(1)_H$ as $1/2$. The SM matter hypercharges are
\begin{align}
  \label{eq:cp2y}
  Q_Y 
      &= 
  \begin{cases}
    \frac{n}{3},      \quad & \mbox{for integer weak isospin}\\
    \frac{2n + 1}{6}, \quad & \mbox{for half-integer weak isospin}.
  \end{cases}
\end{align}
For the $SU(2)_L$ quark and lepton doublets ($Q$ and $L$), we have $n
= 0, -2$, respectively, to reproduce the correct hypercharges, while
the $SU(2)_L$ singlet electrons, up, and down-type quarks (written as
left-handed fields $\bar{e}, \bar{u},$ and $\bar{d}$) have $n = 3, -2,
1$, respectively.

Once the $SU(2)_L \times U(1)_Y$ are gauged, the NGB is a pseudo-NGB,
gaining a mass from gauge interactions, of order
$\sqrt{\alpha_\mathrm{EW}}\Lambda$, with $\Lambda$ a cutoff and
$\alpha_\mathrm{EW}$ the electroweak strength. We consider $\Lambda
\sim M_p$, the Planck scale, and we need a fine tuning to explain a
light Higgs mass at the electroweak scale.\footnote{This tuning and
  cancellation of loop corrections at the Planck scale is worse than
  the usual considerations (there is no $G/H$-symmetric counter term),
  but the model is consistent once this is imposed. For now we can
  only assume some ``miracle'' at the Planck scale.}  A quartic
self-coupling is generated via gauge and Yukawa interactions at the
Planck scale which is one-loop suppressed and thus sufficiently small,
$\mathcal{O}(10^{-3})$. There may be a connection with models
(e.g.~\cite{Shaposhnikov:2009pv, *Bezrukov:2012sa}) exploiting the
possibility of the Higgs quartic coupling running to zero (within
sizable errors, especially from the top mass) near the Planck scale
(see \cite{0lambda-ref4} and references therein). A NGB hypothesis for
the Higgs may be consistent if there is such a flat potential at the
Planck scale, with appropriate assumptions on UV effects or boundary
conditions. This would be a remarkable observation, connecting the
$126$ GeV Higgs mass \cite{Higgs-ATLAS, *Higgs-CMS} with physics at
the Planck scale and charge quantization.

We also consider the phenomenology of the \cp3 model, where the SM is
now $\cp3 \times SU(2)_L$. The unbroken group from \cp3 is $SU(3)_H
\times U(1)_H$, which we take to be color and hypercharge,
respectively. The proportionality constant between $U(1)_Y$ and
$U(1)_H$ is $2/3$:
\begin{equation}
  \label{eq:cp3y}
  Q_Y = \frac{2}{3}\qHOne = \frac{3n + T}{6},
\end{equation}
where $T$ is the ``$3$-ality'' of the $SU(3)_H$ representation, given
by $T = 0, -1, +1$ for a singlet, anti-fundamental, and fundamental
representation, respectively. This now corresponds to the QCD
representation for the given matter field. The left-handed quark
$SU(2)_L$ doublet $Q$ has $n = 0$ ($T = +1, Q_Y = 1/6$), a down-type
$SU(2)_L$ singlet quark $\bar{d}$ has $n = +1$ ($T = -1, Q_Y = 1/3$),
a lepton doublet $L$ has $n = -1$ ($T = 0, Q_Y = -1/2$), and so on.

In this model we consider the (conjugate) NGB in the anti-fundamental
representation of $SU(3)_\mathrm{QCD}$ with hypercharge $Q_Y = -2/3$:
it has the quantum numbers of an $SU(2)_L$ singlet up squark. In fact,
in a supersymmetric model the partner fermion $\bar{u}$ can explain
the smallness of the up quark mass in the SM. It can even be possible
to have a massless up quark, avoiding the strong CP problem.

\subsection{The SM generation content from \cp2 and \cp3}
\label{sec:sm-cp2}

There is another intriguing phenomenological consequence of the \cp2
and \cp3 models: with some minor assumptions we can obtain the
structure of the matter content of a generation in the SM. We impose
the following restrictions:
\begin{itemize}
  \item There is no vector-like matter (which would then have a natural
    mass scale of $M_p$).
  \item The theory is anomaly free for all gauge groups.\footnote{This
      is reminiscent of an alternative observation of charge
      quantization by examining the SM anomalies (see
      \cite{Foot:1990uf, *Foot:1992ui, *fukugitayanagidabook}). Here,
      however, it is ``opposite'' in the sense that the charge
      quantization rule derived above leads to the SM matter
      content. For the relation of the matter representations and
      anomalies in the SM, see the earlier work of
      \cite{Geng:1988pr}.}
  \item The smallest representations and least amount of matter should
    be used.
\end{itemize}

We will start by looking at the \cp2 model.  We will add the color group, $SU(3)_\mathrm{QCD}$, and one $SU(2)_L$
doublet quark (fundamental of $SU(3)_\mathrm{QCD}$) with hypercharge
\begin{equation}
q_Q = Y\frac{2n_Q + 1}{3},
\end{equation}
where $Y$ is the constant of proportionality between hypercharge and
the NLSM $U(1)_H$ (or equivalently, the NGB charge which all charges
are proportional to), which we will leave arbitrary in this analysis.

Let us first consider the $SU(3)_\mathrm{QCD}^3$ anomaly. Since we have
one quark doublet, based on our assumptions of no vector-like matter
and using minimal content, we must add two $SU(2)_L$ singlet
anti-quarks, with charges $2Yn_{\bar{u},\bar{d}}/3$. The charges in
the $SU(3)_\mathrm{QCD}^2U(1)_Y$ anomaly (all fields are left-handed)
then require that
\begin{equation}
  \label{eq:su33}
  2n_Q + 1 = -(n_{\bar{u}} + n_{\bar{d}}).
\end{equation}

Next,\footnote{Remember that the $SU(2)_L^3$ anomaly, and anomalies
  with one $SU(3)_\mathrm{QCD}$ or $SU(2)_L$ factor, are automatically
  satisfied due to the group structure.} we have the $SU(2)_L^2U(1)_Y$
anomaly. Only the quark doublet contributes, so we must add a lepton
doublet. Writing its charge as $Y(2n_L + 1)/3$ the restriction on the
charges from the anomaly is
\begin{equation}
  \label{eq:lep}
  2n_L + 1 = -3(2n_Q + 1) = 3(n_{\bar{u}} + n_{\bar{d}}).
\end{equation}

Now consider the $(\mathrm{gravity})^2U(1)_Y$ anomaly. Again, we are
required to add additional matter, which will be a singlet except for
its $U(1)_Y$ charge: an $SU(2)_L$ singlet lepton with charge
$2Yn_{\bar{e}}/3$. The anomaly constraint is
\begin{equation}
  \label{eq:e}
  n_{\bar{e}} = -3(n_{\bar{u}} + n_{\bar{d}}).
\end{equation}

Finally, we have the $U(1)_Y^3$ anomaly. There is no extra matter that
is required, if the integers giving the charges satisfy
\begin{equation}
  \label{eq:ud_cond}
  (n_{\bar{u}} + n_{\bar{d}})(2n_{\bar{u}} + n_{\bar{d}})(n_{\bar{u}} + 2n_{\bar{d}}) = 0.
\end{equation}
Combined with the relation to $n_Q$ in eq.~\eqref{eq:su33}, and up to
exchanging $n_{\bar{u}}, n_{\bar{d}}$, the unique solution is
\begin{equation}
  \label{eq:cp2cqsoln}
  n_{\bar{u}} = -2n_{\bar{d}}.
\end{equation}

We have therefore ``derived'' the matter content of a generation in
the SM, with the final form of the $U(1)_Y$ charges given in Table
\ref{tab:cp2_charges}. We still need to fix the overall coefficient,
$Y$, which we will set by taking the Higgs (NGB) hypercharge to be
$1/2$. These charges are of course consistent with the usual SM charge
assignments, as given previously. It seems rather unexpected, and
remarkable, that such structure comes from the \cp2 NLSM with rather
minimal additional assumptions.

\begin{table}
  \centering
  \begin{tabular}{c|c}
    Field & $U(1)_Y$ Charge\\
    \hline
    $H$ & $Y \Rightarrow Y = \half$\\
    $Q$ & $\frac{2n_Q + 1}{6}, 2n_Q + 1 = -(n_{\bar{u}} + n_{\bar{d}})$\\
    $\bar{u}$ & $\frac{n_u}{3}, n_{\bar{u}} = -2n_{\bar{d}}$\\
    $\bar{d}$ & $\frac{n_{\bar{d}}}{3}$\\
    $L$ & $\frac{2n_L + 1}{6}, 2n_L + 1 = 3(n_{\bar{u}} + n_{\bar{d}})$\\
    $\bar{e}$ & $\frac{n_{\bar{e}}}{3}, n_e = -3(n_{\bar{u}} + n_{\bar{d}})$
  \end{tabular}
  \caption{The constraints on the matter and its $U(1)_Y$ charges due to anomaly cancellation requirements in the \cp2 model, with the Higgs $H$ realized as the
  Nambu-Goldstone boson.  The normalization of the hypercharge is determined by the assigning $H$ to have $Y=1/2$. $Q$ and $L$ are the $SU(2)_L$ quark and lepton doublets, while $\bar{u}, \bar{d},$ and $\bar{e}$ are the $SU(2)_L$ singlet up quark, down quark, and electron, respectively. All fermions are written as left-handed fields. The SM has $n_Q = 0, n_{\bar{u}} = -2, n_{\bar{d}} = 1, n_L = -2, n_{\bar{e}} = 3$.}
  \label{tab:cp2_charges}
\end{table}

We can follow basically the same procedure in a supersymmetric \cp3
model. In this case we add the weak group to complete the SM gauge
groups. For this model we will work with an \emph{anti}-fundamental
NGB; it has $U(1)_H$ charge $-1$ and hypercharge $-Y$ (the
normalization to hypercharge). With supersymmetry, the fermion partner
to the NGB is an $SU(2)_L$ singlet quark. Whether it is an up or down
quark depends on how we fix $Y$, which we again leave as a free
parameter at this stage. We write this fermion as a left-handed field
$\bar{u}$ which is an anti-fundamental of color with hypercharge $-Y$.

With just a single colored fermion, we need to add matter to satisfy
the $SU(3)_\mathrm{QCD}^3$ anomaly. If we try adding another weak
singlet, but with a different hypercharge to forbid a mass term, we
then cannot satisfy the $SU(3)_\mathrm{QCD}^2U(1)_Y$ anomaly without
additional colored matter (changing the $SU(3)_\mathrm{QCD}^3$
anomaly). Instead, we add an $SU(2)_L$ quark doublet, with hypercharge
$(3n_Q + 1)Y/4$, and an $SU(2)_L$ singlet (anti-fundamental) quark,
with hypercharge $(3n_{\bar{q}} - 1)Y/4$. The
$SU(3)_\mathrm{QCD}^2U(1)_Y$ anomaly relates their charges as
\begin{equation}
  \label{eq:cp3su3anom}
  2n_Q + n_{\bar{q}} = 1.
\end{equation}

The $SU(2)_L^2U(1)_Y$ anomaly requires a (colorless) weak doublet: the
lepton doublet with hypercharge $3Yn_L/4$. This charge is related to
the quark doublet by
\begin{equation}
  \label{eq:cp3su2anom}
  n_L = -(3n_Q + 1).
\end{equation}

Next, we have the $(\mathrm{gravity})^2U(1)_Y$ anomaly. The anomaly is
zero only with additional matter (as $n_Q \in \mathbb{Z}$): an
$SU(2)_L$ singlet lepton with hypercharge $3Yn_{\bar{e}}/4$. This
charge is related to the quark charges by
\begin{equation}
  \label{eq:cp3gravanom}
  n_{\bar{e}} = 2(3n_Q + 1).
\end{equation}

The final anomaly is the $U(1)_Y^3$ anomaly. For this anomaly to be
satisfied we must satisfy the constraint
\begin{equation}
  \label{eq:cp3u3anom}
  n_Q (n_Q + 1) (3n_Q + 1) = 0. 
\end{equation}
There are two possible unique solutions for the integer $n_Q$ which
specifies all of the matter hypercharges (the Higgs does not appear in
the anomaly constraints). If $n_Q = -1$, then we can use one of the
hypercharges of the SM to fix $Y = -1/3$. All of the hypercharges are
specified and match their SM values, and the partner of the NGB is a
down-type quark.

The more interesting possibility is if $n_Q = 0$. Fixing one of the
hypercharges to the SM value requires $Y = 2/3$, which matches the
``minimality'' considerations we used in the previous section. In this
case the NGB fermionic partner is the up quark. This raises the
possibility of connecting the \cp3 model of charge quantization to the
smallness of the up quark mass and the strong CP problem, which can be
avoided with a massless up quark.


\section{Discussion and Conclusions}
\label{sec:conc}

In this work we have shown how to extend the earlier results for the
\cp1 model \cite{cp1paper} to general \cpk models. The simplest way to
do this is to add mass terms to additional NGBs and flow through
renormalization to the \cp1 model. We then arrive at a general charge
quantization formula, which depends on a matter field's representation
under the unbroken group. We explored some of the phenomenological
implications of these NLSMs with some part of the SM as the unbroken
group.

The \cp2 and \cp3 models have very interesting phenomenology. The NGBs
in the \cp2 model have the quantum numbers of the SM Higgs boson,
which presents some interesting model-building possibilities. In the
\cp3 model with supersymmetry, the fermion partner to the NGB is the
up quark, connecting the model to the possibility of a vanishing up
quark mass as a solution to the strong CP problem. Quite unexpectedly,
both of these models, with some assumptions like chiral matter, lead
to the structure of the SM matter generation content. This is due to
the charge quantization formula enforcing that non-singlet fields have
a nonzero hypercharge.

A logical continuation of this program would be to try to embed the
entire SM as the unbroken group of a NLSM. This is currently under
investigation, to appear in a future work. The charge quantization
formula in this model can again be obtained by considering embedding
the \cp1 model, while the phenomenology is quite rich.

There are also several open questions related to these types of models
which we are currently exploring. One question regards explicit
breaking, beyond that of the gauging of the unbroken group. While the
breaking due to gauging a symmetry is under control, determined by the
(small) gauge coupling, what about other possible sources? Charge
quantization in these models can be thought of as a topological effect
(the structure and compactness of the group manifold), and thus may be
robust against other breaking effects. 
Finally, there are also several interesting topics which are related
to these types of theories which we are exploring. This includes
anomaly considerations, beta functions, and more mathematical
considerations. These NLSMs are proving to have quite a rich
structure, probing deep questions in particle physics and the SM.

\begin{acknowledgments}
  \noindent
  The authors would like to thank Mikhail Shaposhnikov for interesting
  discussions. This work was supported by the World Premier
  International Research Center Initiative (WPI Initiative), MEXT,
  Japan. The work of S.H.~was also supported in part by a Grant-in-Aid
  for Scientific Research (22740153) from the Japan Society for
  Promotion of Science (JSPS).
\end{acknowledgments}

\bibliography{cpk-cq-refs}

\end{document}